\documentclass[12pt,preprint]{aastex}

\newcommand{\lta}{{\>\rlap{\raise2pt\hbox{$<$}}\lower3pt\hbox{$\sim$}\>}}
\newcommand{\gta}{{\>\rlap{\raise2pt\hbox{$>$}}\lower3pt\hbox{$\sim$}\>}}

\shorttitle{Observing the Reionization Sources}
\shortauthors{Stiavelli et al.}

\begin{document}


\title{Observable Properties of Cosmological Reionization Sources}


\author{M. Stiavelli, S. Michael Fall,  and N. Panagia\altaffilmark{1}}
\affil{Space Telescope Science Institute, 3700 San Martin Dr., Baltimore, MD
21218}


\altaffiltext{1}{ESA Space Telescope Division}


\begin{abstract}
Motivated by recent evidence that the epoch of reionization of hydrogen
may have ended at a redshift as low as $z\approx 6$, we consider the detectability of 
the sources responsible for this reionization. The main idea is that
reionization places limits on the mean surface brightness of the population
of reionization sources.
Reducing the number of model-dependent 
assumptions to a minimum, we discuss the observability of 
these sources with existing and planned telescopes. We define a
family of models characterized by two parameters: the Lyman continuum escape fraction
$f_c$ from the sources, and the clumpiness parameter $C$ of the intergalactic medium. 
The minimum surface brightness model corresponds to a value of
unity for both parameters. We find that
the detection of the non-ionizing UV continuum of the reionization sources will be difficult
to accomplish before the launch of JWST if these sources have a mean surface brightness close to 
the minimum value. However, if the values of $f_c$ and $C$ are more realistic, the reionization
sources may well be detected by HST before JWST, perhaps by exploiting gravitational lensing 
amplification by foreground clusters of galaxies.
Instead of a detection in the continuum, one may attempt to detect the 
Ly$\alpha$  emission line by narrow-band imaging.
Present Ly$\alpha$ searches at $z>6$ suggest that either
the typical sources are fainter than dwarf galaxies, or that the escape fraction of ionizing 
photons is much higher than 50\%, so that there are no bright compact HII regions formed 
around the ionizing sources. 
\end{abstract}




\section{Introduction}

The recent discovery of quasars at $z \gta 6$ in the Sloan Digital Sky Survey 
\citep{fan01} has 
given a new impetus to the study of the epoch of reionization.
In particular, SDSSp J103027.10 +055255.0 at $z=6.28$  shows
a  significant continuum absorption shortwards of Ly$\alpha$,
reminiscent of a Gunn-Peterson (1965) trough \citep{becker01,fan02}, and suggests that 
we may be close to identifying -- or may have already identified -- 
the end  of the epoch of hydrogen reionization.
A high priority now is to
identify the sources responsible for producing the UV ionizing radiation 
(the reionization sources)
and to determine their nature. There have been several attempts to predict the
properties of these sources \citep{madau99,barkanaloeb00,miraldaetal00,haimanetal01,ohetal01,brommclarke02,yan02,cen03,fukugitakawasaki03,
venkatesanetal03,wyitheloeb03a,wyitheloeb03b} and several reviews have been published recently 
\citep{loeb02,stia02}. 

Here, we introduce a new approach to explore the observable properties of
the reionization sources, while minimizing the number of model-dependent
assumptions. Our main interest is to determine the minimum cumulative surface brightness (in flux, maximum in 
magnitudes) of the population of
reionization sources rather than to provide a complete
evolutionary model of these sources.
A strict lower limit to the surface brightness of the sources is
obtained by requiring that they produce as many ionizing photons as there are
hydrogen atoms in the universe. This limit can be improved upon
by considering the main physical effects taking place during
reionization and their relative importance. The required surface brightness is minimum when the escape of ionizing
UV radiation is the highest, so that the source luminosity is lowest in the non-ionizing UV and
in the visible. A maximum surface brightness is obtained by requiring that the reionization sources do not overproduce
heavy elements.
Our general approach is applicable to most types of reionization sources, but when specific numerical examples are
given, we illustrate our procedure by focussing on 
population III stars, as these objects have very high effective temperatures and therefore are
very effective producers of ionizing UV photons.
Our mean surface brightness estimates are compared to the parameter space
that can be probed by existing and future telescopes, in order to help 
planning the most effective surveys. 

In Section 2, we derive the minimum mean surface brightness of the reionization sources by
maximizing the production and the release of Lyman continuum photons into the intergalactic 
medium (IGM). We first recover the proportionality between surface brightness of the reionization sources
and the comoving density of hydrogen (2.1). The proportionality constant depends on both the helium fraction
and the spectral energy distribution (hereafter SED) of the reionization sources. We start out with the case of a 
pure-hydrogen IGM (2.2) 
and then consider the effects of primordial helium (2.3). 
The least model-dependent results
in this paper are those for the homogeneous case (2.4), and the corresponding minimum 
metallicity enrichment of the universe (2.5). In Section 3, we consider more realistic models in order
to provide a better comparison with observations. In particular, we discuss the effects of
a limited escape fraction of 
ionizing UV photons (3.1), increased recombinations induced by
density inhomogeneities in the IGM (3.2), a luminosity
function for the reionization sources (3.3), and the detectability of Ly$\alpha$ emission from these
sources (3.4). Finally, 
in Section 4, we discuss
the observability of these sources both from space, either with HST or JWST, and from ground-based
facilities.

\section{The minimum surface brightness for reionization}

\subsection{The surface brightness of reionization sources}

There is a simple connection between the comoving density of hydrogen and the mean surface brightness $J_\nu$ of the
reionization sources in their non-ionizing continua. Neglecting the absorption of non-ionizing photons in the IGM, the surface brightness of a class of sources is simply related to their volume emissivity $E(\nu)$, the power radiated per unit
frequency per unit comoving volume, by 

\begin{equation}
\label{Jnu}
J_\nu = \frac{c}{4 \pi} \int_{t_1}^{t_2} E([1+z(t)]\nu) dt,
\end{equation}
where $t_1$ indicates the cosmic time when the sources turn on and $t_2$ the time when reionization is
complete (see, e.g., Peebles 1993). Here and elsewhere we refer to comoving volumes and densities unless otherwise specified. The production rate of ionizing photons per unit volume $\dot n_c$ is related to the volume emissivity by

\begin{equation}
\label{ngamma}
\dot n_c = \int_{\nu_{H}}^\infty \frac{E(\nu)}{h\nu} d\nu,
\end{equation}
where $h$ is Planck's constant and $\nu_{H} = 2.467 \times 10^{15}$ Hz is the frequency of photons at the 
hydrogen ionization threshold. Here and in the following we denote with the subscript $c$ quantities referring
to Lyman continuum photons.

For a given SED of reionization sources, there is a direct proportionality relation
between the production rate of ionizing photons $\dot n_c$ and the emissivity $E(\nu_0)$ at some
reference frequency $\nu_o$. With this in mind, we
define the proportionality coefficient

\begin{equation}
\label{defineAofnu}
A(\nu_o) \equiv \frac{E(\nu_o)}{\dot n_c}.
\end{equation}
In the following, we choose $\nu_0$ to be a fixed frequency in the observer frame in the non-ionizing continuum
of the sources. 

In order to ionize hydrogen completely, the total number of ionizing photons per comoving volume must exceed the comoving number density of hydrogen $n_H$ by some factor $B$ that depends on the details of the recombination process and on the helium fraction

\begin{equation}
\label{nhydrogen}
B \equiv \frac{1}{n_H}\int_{t_1}^{t_2} \dot n_c dt.
\end{equation}

Combining equations (\ref{Jnu}), (\ref{defineAofnu}), and (\ref{nhydrogen}), we obtain

%

\begin{equation}
\label{simpleJnu}
J_\nu = \frac{c {\bar A(\nu)} B}{4 \pi} n_H,
\end{equation}
where ${\bar A}(\nu)$ denotes the average of $A(\nu_0)$ weighted by the production rate of ionizing photons, namely

\begin{equation}
\label{defineAterm}
{\bar A(\nu)} \equiv \frac{\int_{t_1}^{t_2} A([1+z(t)]\nu) {\dot n}_c dt}
{\int_{t_1}^{t_2} {\dot n}_c dt}.
\end{equation}
Equation (\ref{simpleJnu}) establishes a simple, direct relation, between the mean surface brightness of reionization sources and the comoving density of hydrogen atoms. In the following, for simplicity, we will 
neglect possible variations in $\dot n_c$. In this case, $\bar A(\nu)$ becomes a time average:

\begin{equation}
\label{simpleAterm}
{\bar A(\nu)} \equiv \frac{1}{t_2-t_1} \int_{t_1}^{t_2} A([1+z(t)]\nu) dt.
\end{equation}

The volume emissivity $E(\nu)$ for a population of sources is given in terms of their luminosity function per comoving
volume, $\Phi(L_\nu)$, as

\begin{equation}
\label{emissivityofPhi}
E(\nu) = \int_0^\infty \Phi(L_\nu) L_\nu dL_\nu.
\end{equation}

Equations (\ref{Jnu}) and (\ref{emissivityofPhi}) establish the connection between the mean surface brightness of a
population of sources and their luminosity function. In the case of identical sources the luminosity function is
simply a delta-function.  The following subsections will be devoted to estimating the values of ${\bar A}(\nu)$ and $B$.

\subsection{Reionization in a hydrogen-only IGM}

Due to recombinations, the minimum number of ionizing photons per comoving volume needed to reionize hydrogen 
exceeds the  number of hydrogen atoms ($B>1$). 
The equation for the ionized fraction $x\equiv n_p/n_H$, including recombinations is

\begin{equation}
\label{ionizedfracoft}
\frac{dx}{dt} = \frac{\dot n_c} {n_H} - C \alpha_B x (1+z(t))^3 n_e,
\end{equation}
where $n_e$ is the comoving density of electrons, $C \equiv <n_p^2>/<n_p>^2 = <n_e^2>/<n_e>^2$ is the IGM clumpiness factor, and $\alpha_B = 1.4 \times 10^{-13}$ cm$^3$ s$^{-1}$
is the recombination rate for an electron temperature of $T_e = 2 \times 10^4$ K \citep{pengelly64}. 
This value of the electron temperature is appropriate for primordial HII regions \citep{daub63, davidsonkinman85, tumlinson01}. The factor $(1+z)^3$ is due to the fact that recombinations depend on the proper 
(not comoving) density of electrons.

For the pure-hydrogen composition adopted in this section we have $n_e = n_p$.
Solving equation (\ref{ionizedfracoft}) and requiring $x(t_2) = 1$, we can determine $\dot n_c$ as a function of $t_1$ and $t_2$. 

The parameter $B$ of equation (\ref{nhydrogen}) is the ratio of the actual number density of ionizing photons to
the number density of hydrogen atoms. We define $B_H$ as the value of $B$ in the pure hydrogen case.
In the absence of recombinations (i.e. in the limit of a very low density) and in the pure hydrogen case, the required 
comoving number density of ionizing photons,
$n_{c, {\rm noHe}} = \dot n_c (t_2-t_1)$,
equals the comoving number density of hydrogen, $n_{c,{\rm noHe}} = n_H$ and therefore $B_H=1$. When recombinations are considered in the homogeneous case ($C=1$),
we find $B_H \leq 1.03$ or $\leq 1.08$ for $z_2 = z(t_2) \lta 10$,  and $\Delta z \equiv z(t_1)-z(t_2) = 1$ or 3, respectively. We also find that if reionization occurs at $z_2=6$, $B_H \leq 1.3$ 
for $\Delta z \leq 50$. Thus, in a homogeneous ($C=1$) pure-hydrogen IGM, recombinations are only a minor 
correction.

\subsection{Reionization in a hydrogen-helium IGM}

More realistically, the chemical composition of the universe at reionization includes both hydrogen and helium, 
with mass fractions $X=0.76$ and
$Y=0.24$ \citep{pagel00}. Thus, the mean molecular weight is given by 
$(m_H Y/m_{He}+X)^{-1} m_H \simeq 1.2 m_H$, where $m_H$ and $m_{He}$ are 
the masses of hydrogen and helium atoms, respectively.
Assuming $\Omega_m = 0.268$, $\Omega_b = 0.044$, $\Omega_\Lambda = 0.732$,
and $H_0 = 71$ km s$^{-1}$ Mpc$^{-1}$ (the best-fit WMAP parameters; Spergel et al. 2003), 
we find that the number density of hydrogen atoms (including ions) per 
comoving volume, $n_H$ is  $2.1 \times 10^{-7} {\rm cm}^{-3}$ or $6.2 \times 10^{66} {\rm Mpc}^{-3}$.

Photons with energies above 24.4 eV can ionize He as well as H. The cross section for He ionization at the
threshold is higher than that of H by about a factor 7 but the number density of He is 
lower by a factor of about 12, so that photons at these energies may be absorbed by, and ionize, either atoms.
In the following we will assume that He absorbs as many photons as possible in proportion to its abundance relative 
to H, i.e. about 8 \% of the total budget of ionizing photons. 
Photons with energy above 54.4 eV can ionize helium twice. 
Whether helium is ionized or not and how many times, depends on the hardness of the UV spectrum. 
As an example, a blackbody at $T=10^5$ K emits 60\%
of its ionizing photons at energies greater than 24.4 eV. In this case, all helium will be ionized, and about
10 \% will be twice ionized. Therefore,
we have $n_e \simeq 1.09 n_p$, and the required minimum number of ionizing photons will be higher by 9\% relative
to the pure H case. Note
that if the SED were hard enough to double-ionize all helium, $n_e$ would increase to $1.16 n_p$. This is unlikely to
happen for stellar sources, but it might for active galactic nuclei (AGNs).

Helium ionization has an additional, indirect effect on the required number of ionizing photons because the 9\% increase in the electron number density makes recombinations slightly more effective. This can be estimated by integrating equation (\ref{ionizedfracoft}) for $n_e \simeq 1.09 n_p$, from which we obtain

\begin{equation}
\label{requiredphotonswithhelium}
B \simeq 1.09 [1+1.09(B_H-1)] B_H.
\end{equation}

In general, we find that primordial helium recombinations 
increase $B$ by less than an additional 1\% for $z_2 \leq10$ and $\Delta z \leq 3$
and by less than 2\% for reionization at $z_2=6$ and $\Delta z \leq 50$. Thus, recombinations remain a minor
correction in the homogeneous ($C=1$) case, even for a hydrogen-helium IGM.

\subsection{Results for a homogeneous IGM}

So far, the only dependence of our calculations on the type of ionizing sources has been in the assumption that all helium is ionized. Indeed, even if this assumption were not verified, our results would change only by 10 \% or less.
In order to derive the mean surface brightness of sources from their production of ionizing photons, we need to assume a specific SED
and compute the quantity ${\bar A(\nu)}$. 
For definiteness, we will now consider population III stars as the sources responsible for reionization of hydrogen. Pop III stars are characterized by virtually zero metallicities and consequently very high effective temperatures \citep{castellani83,tumlinson00,marigo01,bromm02}. Here we will describe them as blackbodies with effective temperatures
of $10^5$ K. Note that high temperatures around $10^5$ K are reached even if the mass function does not extend beyond 
$100 M_\odot$ \citep{tumlinson00,baraffe01,schaerer02}. Pop III stars thus provide us with the minimum mean surface brightness needed for the case of reionization by stars. In fact, for a fixed flux of ionizing photons, cooler star are brighter in the visible and in the non-ionizing UV continuum. One can define the efficiency
of an ionization source as proportional to the ratio of the number of ionizing photons to the total energy output. Cooler stars emit a smaller fraction of their radiation
in the form of ionizing photons and are therefore less efficient ionizers. As an example, for equal ionizing fluxes, a stellar population with $T_*=5 \times 10^4$ K is a factor $\sim$4.5 brighter at $\lambda = 1400$ \AA\ than a stellar population with $T_*=10^5$ K.  A similar comparison can be made for typical AGNs. We find that for typical 
power-law SEDs, AGNs are less efficient ionizers than Pop III stars. 

Once the SED is specified, we can determine the value of $\bar A$ from equation (\ref{defineAofnu}) and (\ref{simpleAterm}). In integrating
equation (\ref{simpleAterm}), the emissivity can be assumed to be zero below the rest-frame wavelength of 1216 \AA\ 
since during the reionization era these photons will be absorbed by intervening neutral hydrogen. 
In magnitude units, the surface brightness per square arcmin 
is given by $\mu_{AB} = -2.5 log J_\nu+8.9$, when $J_\nu$ is expressed in Jy arcmin$^{-2}$.

In Table 1, we give the minimum surface brightness for reionization sources for a variety of terminal redshifts of reionization $z_2$ and durations $\Delta z$. Two values of surface brightness are given in the table
for each combination of $z_2$ and $\Delta z$, corresponding to the rest-frame 1400 \AA\ wavelength observed 
at the terminal redshift of reionization $z_2$ or at the mean redshift $\bar z = (z_1+z_2)/2$. Thus, they are computed at
the observer wavelengths $\lambda = 1400 (1+z_2)$ \AA\ and $\lambda = 1400 (1+ {\bar z})$ \AA\ , respectively.
In addition, we also give the number of photons per ionization $B$ and the Thomson opacity $\tau_T$, integrated to the beginning of the reionization process, and given by

\begin{equation}
\label{Thomson}
\tau_T = \frac{c\sigma_T}{H_0} \int_0^{z_1} \frac{n_e(z)}{(1+z)[\Omega_m (1+z)^3+\Omega_\Lambda]^{1/2}}
dz,
\end{equation}

\noindent
where $\sigma_T = 6.653 \times 10^{-25}$ cm$^2$ is the Thomson cross-section.

All results in Table 1 include the effects of helium ionization (by population III stars) and recombinations. 
It is clear that in the homogeneous case ($C=1$), the minimum surface brightness (in flux, maximum in 
magnitude) is essentially independent of the terminal redshift of reionization (within 0.1 magnitudes for
$\Delta z=1$) but does depend on $\Delta z$.

The value of the Thompson opacity is a strong function of $\Delta z$. In the case of
extended reionization, $\Delta z = 50$, we obtain values of $\tau_T \simeq 0.15$ compatible with the WMAP result 
($\tau_T = 0.17 \pm 0.06$ for the combined fit). The empirical 
2$\sigma$ limit, $\tau_T \geq 0.05$, implies $\Delta z \geq 1$. In the figures, we will focus on the $\Delta z=1$ case
since it is close to the minimum ionizing photon requirement $B=1$ (obtained in the absence of recombinations), while
remaining marginally compatible with the WMAP results.

The numbers given in Table \ref{tbl-1}
can be used directly to obtain the luminosity of the sources of ionizing photons on the assumption that they are a population of identical objects with a specified surface number density. Indeed, the numbers in the table represent the apparent magnitudes of individual objects if their surface number density is 1 arcmin$^{-2}$. A simple rescaling yields the apparent magnitude $m$ for different assumed surface densities $\cal N$ namely

\begin{equation}
\label{formulaNino}
log {\cal N} = 0.4 (m-\mu_{AB}),
\end{equation}
where $\mu_{AB}$ is the mean surface brightness given in Table 1. This relationship for $z_2=6$, $\Delta z=1$, and an observer wavelength $\lambda = 1400 (1+z_2)$ \AA\ is
shown in the left panel of Figure 1 as the solid line labeled $(1,1)$ (this notation will be explained in the next
Section). Models characterized by different values
of mean surface brightness are obtained by simple translations with respect to this reference model.

Note that in the idealized case of identical sources, an entire population of such sources is represented by a single point in diagrams such as Figure 1. Those lying on the indicated line in the diagram would be sufficient to ionize the IGM. Observations that fail to detect sources at some surface density-luminosity
combination along the line would not rule out the model but instead could indicate that the sources lie at some
other point along the line. The shaded area below the line indicates the
area where sources would not produce enough ionizing photons for reionization. 

The results for a different
effective temperature are illustrated in the right panel of Figure 1, which is
analogous to the left panel but refers to sources with $T_*=5 \times 10^4$ K. 


\subsection{Mean metallicity at reionization}

It is interesting to compute the mean metallicity to which the universe is enriched by stellar populations with
the minimum mean surface brightness required to ionize the IGM. Here we define metallicity as $Z=\Omega_Z/\Omega_{b}$, where $\Omega_Z$ is the fractional comoving density in
metals (in both galaxies and the IGM). Massive stars with effective temperatures in the range $3 \times10^4$ to $10^5$ K produce $~10^{62}$ Lyman continuum photons per unit of stellar mass. Assuming a typical yield in metals of 0.2 \citep{woosley95}, we find that the massive stars responsible for reionization enrich the universe to a minimum metallicity of $Z_{min}\sim 1.2 \times 10^{-4}$ Z$_\odot$. This result does not change by more than a factor 3 for a wide range of stellar properties. As an example,
by considering only supermassive Pop III stars, with masses between 100 and 1000 $M_\odot$ with the yields as given
by \citet{heger}, and a power-law mass function with a slope between $-1$ and $-2$, we find that the mean mass-weighted yield is $\sim0.2\pm0.05$ (neglecting the metals trapped in black hole remnants). Stars in this mass range will produce $7.6 \times 10^{61}$ ph $M_\odot^{-1}$ \citep{tumlinson01}, so that the minimum surface brightness model enriches the universe to a mean metallicity  
$Z_{min} \approx 1.7 \times 10^{-4} Z_\odot$. 
For a given IMF, any reduction of the output of ionizing photons can be regarded as an inefficiency in the energetics of reionization
and will increase the fraction of hydrogen that needs to undergo nuclear processing and therefore will increase the mean metallicity. Thus, the maximum efficiency, minimum surface brightness model also corresponds to the minimum metallicity case. 

Estimates of the mean density of heavy elements in the universe at the end of reionization ($z\simeq6$) place additional constraints on the mean surface brightness of the stellar component of reionization sources. It is difficult to determine $\Omega_Z$ precisely since some metals will be in stars, some in the interstellar media of galaxies, and some in the IGM. But we can bracket the likely range of metal densities as follows. \citet{songaila01} has estimated a mean metal density in the IGM of $\Omega_{Z,IGM} \sim 3 \times 10^{-7}$, corresponding to $Z \sim 3 \times 10^{-4} Z_\odot$, at $z\simeq5$ from the statistics of CIV and SiIV absorption-line systems. This provides a firm lower limit to the mean metal density shortly after the end of reionization. This limit is about twice the metallicity of the minimum surface brightness model $Z_{min}$ (discussed above).

The present mean metallicity in galaxies is close to solar \citep{kulkarnifall02} and the present mean density of the stellar and interstellar components of galaxies is $\Omega_{gal} \approx 4 \times 10^{-3}$ \citep{fukugitaetal98}. Together these imply $\Omega_{Z,gal} \sim 10^{-4}$ at $z=0$. A variety of observations and models indicate that the fraction of mass in galaxies $\Omega_{gal}$ was smaller at $z\simeq6$ than at $z=0$ by a factor of 10 or more, and that the mean metallicity in galaxies was also smaller by a similar factor \citep{peietal99}. From these considerations we infer a generous but non-rigorous upper limit on the mean metal density of $\Omega_Z \leq 10^{-5}$, corresponding to $Z_{max} = \Omega_Z/\Omega_b \leq 0.01 Z_\odot$ at $z\simeq6$.
This value is $\sim 60$ times higher than the metallicity of the minimum surface brightness model. 

If metal-producing
sources are surrounded by local HII regions, their luminosity at $\lambda=1400$ \AA\ can be enhanced by up to a factor 3 due to the nebular continuum contribution (see Section 3.1). Including this factor,
we use the limit to metal production as an upper limit to the mean surface brightness of reionization sources. This is obtained for $z_2=6$ and $\Delta z=1$. Reionization at higher redshifts or occurring over a more
extended time interval, produces the same (or higher) metallicity for fainter sources. Thus, the regions under the global metallicity constraint lines in Figure 1 define necessary but not sufficient conditions to guarantee that metals are not overproduced. All reionization models compatible with the global metallicity constraint must lie in the non-shaded region in Figure 1.
Our results are broadly compatible with those of \citet{miraldarees} once the different assumptions
are taken into account.

\section{More realistic models}

\subsection{Escape of ionizing photons}

It is likely that some Lyman continuum photons are absorbed within the reionization sources themselves. Even in an environment with primordial composition, and thus free of dust,
a lower escape fraction $f_c$ can result if the individual sources
are surrounded by an envelope of neutral hydrogen left by incomplete star formation in protogalaxies. In such clouds,
recombinations would occur at much higher rates; these objects would
produce local HII regions for at least part of their lifetimes and thus could display hydrogen and helium emission lines in their spectra \citep{bromm02,haiman02}. We do not know the best value of the mean escape fraction. It only takes a column density of $1.7 \times 10^{17}$ hydrogen atoms cm$^{-2}$ to produce unit optical depth at the Lyman limit ($\lambda=912$ \AA). Moreover, some amounts of dust may well be present because of enrichment by previous generations of stars. 
\citet{steidel01} derive from their observations, corrected for intervening IGM absorption, a reduction of the Lyman continuum with respect to the continuum
at 1500 \AA\ 
by $\sim 50$ \%. They estimate that dust absorption in the far-UV be a factor of $\sim 5$. Thus, the total combined effect of
gas and dust gives $f_c \approx 0.1$. The \citet{steidel01} observations also suggest that metal-free, dust-free galaxies might have $f_c \approx 0.5$. However, the properties of the reionization sources at 
$z>6$, including the Lyman continuum escape fraction, may be different from those of the Lyman-break galaxies at $z\simeq3$.
\citet{leitherer95} argue for escape fractions below 15\% in bright starburst galaxies in the local universe. However, these observational studies measure the escape fraction near 912 \AA\ while the hydrogen ionization cross section decreases rather rapidly with decreasing wavelength so that even with optical depth $\tau = 2-3$ at 912 \AA\ one could have a Lyman continuum escape fraction greater than 20 \%, especially for supermassive stars that are hotter than local O stars \citep{mezger78,panagia03}.  There have also been a few attempts to estimate the escape fraction theoretically based on idealized assumptions on the structure of the sources. The derived values of the escape fraction span a relatively wide range roughly centered around $f_c \simeq 0.1$ \citep{doveshull94,doveetal00, ricottishull00}.
In the absence of
a definite value, we will consider two possible escape fractions: $f_c=0.5$ and $f_c=0.1$.

An escape fraction $f_c < 1$ increases the required mean surface brightness relative to the minimum value. This in turn
increases the required value of $B$ by a factor $f_c^{-1}$.
Moreover, the fraction of Lyman-continuum radiation that does not escape generates a nebular continuum (mostly two photon emission)
in the rest-frame non-ionizing UV, which dominates the observed continuum flux at $\lambda=1400$ \AA\ \citep{panagia03}. Combining the two effects, we find that the actual continuum flux $F$ at $\lambda=1400$ \AA\ is higher than the value for the complete escape of ionizing photons by the factor

\begin{equation}
\frac{F(f_c)}{F(1)} \simeq \frac{3-2f_c}{f_c} ,
\end{equation}
where the coefficients 3 and 2 account for the fact that the nebular continuum at 1400 \AA\ for population III HII regions is
roughly twice as high as the stellar continuum.

\subsection{Clumpy IGM}

In addition to inhomogeneities near and within the sources, there will also be density fluctuations on larger scales. Such inhomogeneities will increase the recombination rate and act as sinks of ionizing UV photons. This effect can be included by changing the value of the clumpiness factor $C$ in equation (\ref{ionizedfracoft}). We have considered
the value of $C=30$ \citep{gnedinostriker97,madau99} and recomputed the mean surface brightness and the Thomson optical depth $\tau_T$ for the same models that we considered for $C=1$. The resulting values are also given in Table 1. Based on the simulations by \citet{gnedinostriker97}, the value of $C=30$ appears to be a reasonable choice for $z \simeq 6-7$. At higher redshifts, smaller values would be expected. Thus, it is likely
that the effective value of $C$ is bracketed by the two values ($C=1$ and 30) we consider here.

As a result of gravitational instability, more reionization sources will form and the IGM will become more clumpy as time increases. Thus, the increase in the ionization and recombination rates will at least partially offset each other. This argument suggests that our results, based on the idealization that $\dot n_c$ and $C$ are constant, will 
be more accurate than they might at first appear.


An interpolation formula connecting the parameters $B$ and $C$ is given in the Appendix.
We find that the value of $C$ does not affect the optical depth $\tau_T$ significantly. In contrast, the surface brightness increases by a factor of 2-10 (i.e. 0.8-2.5 magnitudes) for the models with $C=30$, as shown in Table 1.

\subsection{Two-parameter models}

In the following, we denote models with the notation ($f_c$, $C$) to indicate the adopted values of the Lyman-continuum escape fraction and of the clumpiness parameter. The (1, 1) label indicates the minimum surface brightness model as described in the previous section. The $B$ factor of equation (\ref{simpleJnu}) is now a function of $f_c$ and $C$.  Any model with $C=1$ located below the metallicity constraint in Figure 1 is allowed. 
Note that the maximum-metallicity constraint was derived in the limit of very small $f_c$ so that the nebular
continuum can enhance the source luminosity by up to a factor of three. Models with $f_c \simeq 1$ and $C>1$ would have little or no contribution
from nebular continuum but would be intrinsically bright and therefore produce more metals. Thus,
it is possible that some models with $C >1$ and $f_c \simeq 1$ lie below the shaded area but still violate the maximum metallicity constraint. In Figure 1, we plot as thin dotted lines the luminosity-surface density relations for identical reionization sources from models with $(f_c,C)=(0.5, 1)$, and $(0.1, 30)$. The lines were computed for $z_2 =6$ and 
$\Delta z= 1$ but are nearly identical for all $z_2 < 10$. 
The mean surface brightness of the reionization sources changes with $\Delta z$ by less than a factor 4 for all $C\leq30$
(see Table 1). For the sake of completeness, we have explored a wider range of parameters than it is probably realistic.
Note that both of the (0.5,1) and (0.1,30) models are allowed 
by the maximum metallicity constraint.

\subsection{Luminosity function of ionizing sources}

So far, we have considered reionization sources with identical luminosity, but it is more likely  that they are characterized
by a broad luminosity function. On the basis of both theoretical considerations and  experience at lower redshifts, it is reasonable to adopt a \citet{schechter} differential luminosity function, $\Phi(L) \propto L^{-\alpha} \exp {(-L/L_*)}$. Here, we adopt the parameters derived for Lyman-break galaxies at $z\simeq3-4$, i.e.  $\alpha=1.6$ and 
$M_{*,1400} = -21.2$
as our reference luminosity function \citep{steidel99,yan02} and explore the effects of variations in both $M_{*,1400}$ and $\alpha$. 
The top left panel of Figure 2 was computed for the reference cumulative
luminosity function in the case of $z_2=6$ and
$\Delta z=1$. The bottom solid line
refers to the minimum surface brightness (1,1) model. 
The area beneath this line
would not reionize the IGM. The top solid line shows the metallicity constraints and the shaded
area above this line shows the region where -- with the adopted luminosity function -- metals would be overproduced. The (0.5,1) and (0.1,30) models are shown as dotted lines.
The remaining panels of Figure 2 explore the effects of varying $M_{*,1400}$. The top right panel
is for $M_{*,1400}$ = -23.2, i.e. two magnitudes of luminosity evolution compared to $z=3$ Lyman break galaxies.
The bottom left panel is for $M_{*,1400}$ = -17.5 corresponding to
a star cluster with $10^6M_\odot$ in massive stars forming over a period of $\sim 1$ Myr. Finally, the bottom right
panel is for $M_{*,1400}=-13.5$ which represents 7.7 magnitudes of luminosity evolution compared to the $z=3$ Lyman break galaxies and corresponds to the case of small dwarf galaxies or proto-globular clusters. The non-shaded area in the figures indicates the allowed region for each luminosity function.
The curves would shift by less than 0.1 magnitudes for any value of $z_2 < 10$. 

In contrast to those in Figure 1, the lines in Figure 2 represent luminosity functions that need to be populated
everywhere in order to produce the required ionizing flux. Thus, a survey ruling out sources at one point along
a curve would rule out that particular combination of $(f_c, C)$ model and luminosity function.

In Figure 3, we explore the effects of changing the slope of the luminosity function. The top left panel is identical to that
of Figure 2 and is repeated here for convenience. The top right panel is for $\alpha=1.1$, similar to the slope of the $z=0$ luminosity function. The bottom left panel is for $\alpha=1.9$.
The bottom right panel of Figure 3 shows the forbidden region (double shading) and the preferred region (no shading) 
for a range of models. In this
last panel, we have considered models with $M_{*,1400}$ ranging from -21.2 to -16.2 and luminosity function slopes
ranging from $\alpha=1.1$ to 1.9. The lower shaded area represents the region of underproduction of ionizing
photons common to all these luminosity functions. Similarly, the upper shaded area corresponds to the common region of overproduction of metals. The non-shaded area is allowed for all the models within the adopted range of parameters.
Thus, this area should be favored in searches for the reionization sources.

The analogs of Figures 2 and 3 for models other than those with $z_2=6$ and $\Delta z=1$ need in general to be recomputed by properly renormalizing the luminosity function. However, these curves can be obtained approximately
by translating vertically the curves for the $(1,1)$ model with $z=6$ and $\Delta z=1$ by an amount $\Delta log {\cal N}
= -0.4\Delta \mu_{AB}$, where $\Delta \mu_{AB}$ is the difference between the values of mean surface brightness of the models considered. This procedure is only approximate because it does not properly average the value of $M_{*,1400}$ 
over the redshift interval $\Delta z$ being considered.

\subsection{Detecting Ly$\alpha$ from ionizing sources}

The fraction $1-f_c$ of the ionizing continuum that does not escape generates local HII regions around the reionization sources, making them potentially visible in the H recombination lines and, more faintly, in the HeI lines.
Prior to full reionization, when the universe consisted of sources embedded in their own HII regions, Ly$\alpha$ photons were able to escape if
these HII regions were big enough that their outer boundaries were redshifted out of resonance 
\citep{madau00,haiman02}.   In the absence of dust, the fraction of Ly$\alpha$ photons that escape from the HII region is determined solely by its size and hence by the production rate of ionizing photons of its central source
$\dot N_c$  \citep{loebrybicki}. 
In the more realistic case, where some dust is present, an even smaller fraction of Ly$\alpha$ photons would escape
\citep{panagiaranieri73,bonilhaetal79,fallcharlot}. Neglecting absorption by the neutral medium, the Ly$\alpha$ line intensity can be computed as follows. Each recombination within the HII region has a 69\% probability of producing one decay leading to Ly$\alpha$ emission. Under conditions of very low metallicity ($Z<10^{-2} Z_\odot$), the Ly$\alpha$ production
is increased by $\sim$50 \% due to collisional excitation \citep{panagia03}. The number density of ionizing photons 
absorbed within the local HII region is obtained from the required number density of photons needed to reionize the universe, $n_c$ as $(f_c^{-1}-1) n_c$, where $f_c$ is again the escape fraction of Lyman continuum photons. Following \citet{miraldarees} -- and in analogy to equation (\ref{simpleJnu}) -- we find that the rate of production of Ly$\alpha$ photons per unit solid angle is:

\begin{equation}
\frac{d{\dot N}_\alpha}{d\Omega} = \frac{c}{4 \pi} (f_c^{-1}-1) n_c.
\end{equation}
The Ly$\alpha$ surface brightness is the product of $d{\dot N_\alpha}/d\Omega$ and the mean observer frame 
photon energy $h {\bar \nu}_{\alpha}$.
For $\Delta z=1 $ and $z_2=6$, this surface brightness is  $(f_c^{-1}-1) \times 1.276 \times 10^{-16}$ erg cm$^{-2}$ s$^{-1}$ arcmin$^{-2}$. This value is a strong function of $C$, decreases with increasing $z_2$,
and depends only slightly on the duration of reionization. We now define $f_\alpha$ as the escape fraction
of Ly$\alpha$ photons. The  average surface brightness in Ly$\alpha$, $J_\alpha$, of the reionization sources corresponding to the  $(f_c, C)$ model is then expressed as

\begin{equation}
J_{\alpha} = \frac{c}{4 \pi} B (f_c^{-1}-1) f_\alpha h {\bar \nu}_{\alpha} n_H.
\end{equation}

\noindent
We have computed
$f_\alpha$ following \citet{miralda98} and \citet{madau00}, assuming a line width of 150 km s$^{-1}$. 
In the left panel of Figure 4, we show the locus of identical Ly$\alpha$ sources corresponding to $(f_c,C)=(0.5, 1)$ and $(0.1,30)$ and for $z_2=6$ and $\Delta z=1$. We also show the metallicity constraint as the upper shaded area. In the
case of Ly$\alpha$ sources, there is no lower shaded area, indicating insufficient production of ionizing photons,
as sources with $f_c$ close to unity have arbitrarily low Ly$\alpha$ flux but are still capable of ionizing the IGM. The right panel of Figure 4 shows two luminosity functions for the Ly$\alpha$ sources. The solid line represents the same luminosity
function as the $z=3$ Lyman-break galaxies converted to Ly$\alpha$ luminosity for the (0.5,1) model. The dashed
line represents a luminosity function with $\alpha=-1.1$ and $M_{*,1400}=-17.5$.

In addition to the case of an intrinsic width of 150 km s$^{-1}$, we have also considered cases with 75 to 300 km s$^{-1}$. As expected, the wider the line,
the more Ly$\alpha$ radiation escapes. 
For example, for sources at the faint end of Figure 4, the fraction of escaping Ly$\alpha$ photons is $\sim 30$ \% higher for $v=200$ km s$^{-1}$ and $\sim 30$\% lower for $v=100$ km s$^{-1}$.

\section{Present and Future Surveys}

Our models can be used to plan and interpret searches for reionization sources with present and future telescopes.
A possible approach to identifying the reionization sources is the Lyman-break technique, which consists
of broad-band photometry with three filters, two to measure the continuum longwards of Ly$\alpha$ and a third filter to measure  (or place upper limits on) a much lower flux just below Ly$\alpha$. Note that for
$z \lta 4$, the
technique places the shortest wavelength filters across the Lyman-continuum break, while at $z\gta 5$ the combined effect of the Ly$\alpha$ forest shifts the wavelength of the break to Ly$\alpha$.
In Figures 1, 2 and 3, we identify the region in the surface density vs luminosity plane probed by the GOODS ACS survey \citep{Giavalisco}, the HDF \citep{Thompson} and HDFS \citep{WilliamsHDFS} NICMOS fields, the Hubble Ultra Deep Field (UDF) and a hypothetical survey with the James Webb Space Telescope (JWST) with 100 hr of integration per filter. 
The limiting magnitudes are those corresponding to a 10$\sigma$ detection in the prescribed integration time.
The limiting surface densities of these surveys has been determined by requiring that at least three sources be detected at or above 10$\sigma$ significance level in the survey area. Thus, the limiting
surface density is given by ${\cal N}_{lim} = 3 / ({\rm survey\ area})$.
These two limits define the quadrant in the $log {\cal N} - m_{AB}$ plane where a particular survey is sensitive to
reionization sources, i.e. higher $\cal N$ and brighter magnitudes. Note that the curves in Figure 1 again apply to
the case of identical sources. Thus, a survey not finding sources at its limiting depth and surface density does not rule out
a particular model but only a particular luminosity-surface density combination. In contrast, in Figures 2 and 3 where
broad luminosity functions are plotted,
all sources along the particular curve being considered are needed to reionize the IGM. Thus,
a survey not finding sources at its limiting luminosity and surface density can rule out that model-luminosity
function combination.

It is clear that GOODS/ACS observations cannot reach the faintest predictions shown in the figures. 
Depending on the actual luminosity function parameters $\alpha$ and $M_{*,1400}$, GOODS/ACS may be more or less successful than the UDF in identifying
$z=6$ reionization sources.
In general, ACS observations could identify reionization sources with the F850LP filter if reionization occurred at a low redshift (z$\simeq$6) and if the reionization sources had a luminosity function
similar to that of Lyman-break galaxies. However, even for $z_2=6$, one can detect all reionization sources as F775W dropouts only
if $\Delta z \lta 0.5$. For larger values of $\Delta z$ or higher values of the reionization redshift, many sources would drop out of F850LP and could be identified only by a deep follow-up study, e.g., with the IR channel of WFC3 \citep{stiavelliISR}. While the HDF/HDFS NICMOS observations could be used to probe
higher redshifts, they lack the combination of area and sensitivity necessary to detect reionization sources as faint as the minimum surface brightness ones. In contrast, JWST has both the wavelength coverage and the sensitivity to detect minimum surface brightness reionization sources for all the models discussed here, provided that $M_{*,1400} < -14.2$; see, for example, the bottom panels in Figure 2.

Another approach to search for the reionization sources
before the advent of JWST is to use gravitational lensing by foreground clusters of galaxies to increase the sensitivity of the surveys \citep{ellis01,hu02}. In Figure 5, the line identified as "Lens" represents a hypothetical survey with 100 orbits per filter with the ACS. The field is centered on one cluster of galaxies at $z=0.5$, idealized as a singular isothermal sphere with a one-dimensional velocity dispersion of 1000 
km s$^{-1}$, corresponding to $T_X \simeq 10$ keV. The faint-end of the line represents an amplification by a factor 10. 
The slope of this line in the intrinsic density of sources (in the source plane) vs. flux plane is -2. This slope corresponds to a constant observed surface brightness since the apparent ({\it i.e.} in the lens-plane) density of sources is lower by a factor identical to the gravitational amplification.
The luminosity function assumed for this figure has a slope $\alpha=-1.6$ and a knee at $M_{*,1400} = -17.5$. Clearly, these sources would be too faint to detect in a survey like the UDF but would be detectable in a 400
orbits survey with ACS pointed toward a suitably chosen cluster of galaxies acting as a lens.
Such a survey could probe an interesting area of parameter space, especially if followed-up by WFC3 near-IR imaging to extend its sensitivity to higher redshifts.

An alternative approach to the Lyman-break technique is the direct detection of Ly$\alpha$ emission. As pointed out by \citet{haiman02} and others and discussed in Section 3, Ly$\alpha$ emission can be detected, with less sensitivity, even from sources at redshifts before reionization is completed (i.e., $z>z_2$)
as long as either the line is intrinsically wide or the source is bright enough to generate its own "proximity effect" by ionizing neutral hydrogen in its neighborhood. Figure 4 shows the detection limits for a survey based on the narrow-band excess technique \citep{rhoads01,hu02}, which consists in obtaining narrow-band exposures to detect line emission and broad-band images or adjacent narrow-band exposures to measure the continuum level. We assume that the survey would be carried out at $z>6$ using a large field-of-view camera on a 4-meter ground-based telescope. In the case illustrated in Figure 4, the search is assumed to cover $\sim 0.6$ square degrees with $\Delta \lambda / \lambda \simeq 0.01$ (labelled LALA in the Figure). Since Figure 4 shows the density of all Ly$\alpha$ sources, the effective area explored by a narrow-band excess survey is reduced by a factor proportional to the $\Delta z$ spanned by the filter.
The oblique marker identified as Lens in Figure 4 represents the detection limits of a 100-orbit ACS grism survey
exploiting gravitational lensing.
In addition, we plot in Figure 4 the upper limits (arrows) and the single detection
(solid vertical bar) reported by \citet{hu02}. Taken at face value, these 
upper limits would rule out, for the present model, values
of $M_*$ brighter than $\sim -17.5$, thus suggesting that only dwarf galaxies exist before reionization. The estimated density of objects like the one detected by \citet{hu02} at $z\simeq 6.56$, by gravitational lensing amplification, would
be sufficient to reionize the universe if these objects had hot ionizing continua with a high escape fraction, $0.1 \leq f_c \leq 0.5$. Clearly, before strong conclusion can be drawn, we need to identify and study more sources
at $z>6$. 

\section{Conclusions}

We have developed a method to predict the observability of cosmological reionization sources based on the fact that
reionization requires a known comoving density of  ionizing photons. This number needs to be at least one per atom,
thus providing a model-insensitive minimum mean surface brightness (in flux, maximum in magnitudes) for the reionization sources. 
This minimum mean surface brightness of reionization sources is general and robust. Our particular choice of cosmological parameters does not affect our results by more than a few tenths of a magnitude.
We find that if reionization is caused by UV-efficient, minimum surface brightness sources, the non-ionizing continuum emission from reionization
sources will be difficult to detect before the advent of JWST. However, the 
actual mean surface brightness may be many times higher than this 
ideal limit if some UV photons are absorbed within the sources 
themselves and/or if some ionizations are offset by recombinations 
in the clumpy intergalactic medium. In this more realistic case, HST 
has a reasonable chance of detecting the reionization sources in 
very deep images, especially if the sources are observed behind clusters of galaxies, so that 
they are magnified gravitationally.
If instead the sources of reionization were not extremely hot population III stars but cooler population II stars or AGNs, they would be brighter by 1-2 magnitudes and 
thus they would be easier to detect.  

The predictions for Ly$\alpha$ emission intensities are less secure than our continuum fluxes because they depend on 
geometry and line width and could be diminished by absorption by
dust, which we have ignored here for simplicity. However, if the Ly$\alpha$ intensities are close to our estimates, searches based on narrow-band excess techniques or slitless grisms would be promising and might lead to the detection of the reionization sources within this decade.



\acknowledgments

We wish to thank Tommaso Treu and Mike Gladders for helpful discussions. Comments from an
anonymous referee contributed to significantly improving the presentation of our results.
This work is partially supported by NASA Grant NAG5-12458.

\appendix
\section{An interpolation formula for $B(C)$}

As discussed in Sections 2 and 3 (see also Table 1) the factor B, which
provides a measure of the extra ionizations needed to compensate for recombinations
and for competing He ionizations, is a sensitive function of both the
clumpiness factor $C$  and the redshift interval $\Delta z$ over which
the universe reionization takes place.  This is because both a high-redshift
and a large $C$ imply higher densities and, therefore, an overall higher
recombination rate that requires a larger number of ionizing photons.

On the other hand, the direct dependence of B on the reionization
redshift  for any fixed $\Delta z$ is much weaker, because when the
reionization process is assumed to take place over a fixed $\Delta z$
interval, for each redshift there is partial compensation between a
higher density $\propto (1+z)^3$ and a shorter duration of the
phenomenon, $\Delta t \propto (1+z)^{-5/2}$.

These facts are clearly illustrated in Figure 6, in which the
behaviour of B as a function of C is shown for
$z_{2}=6$ and several values of $\Delta z$ (left-hand panel), and
for $\Delta z = 1$ and several values of the reionization redshift
(right-hand panel).

Given the importance of the parameter B in cosmological applications, it is
convenient to devise a simple formula that allows a fairly accurate and
quick calculation of B for any value of $z$, $\Delta z$, and $C$, while
retaining the correct functional dependences in the asymptotic limits.
After a little experimentation, we find the following approximate formulae: 

\begin{equation}
\label{BofCapproximation}
B = 1.09 [1+(\xi C)^{5/4}]^{4/5},
\end{equation}

\begin{equation}
\xi = 0.4/[1+(6/\Delta z)(6/z_{2})^p],
\end{equation}

\begin{equation}
\label{exponentp}
p = 0.30+0.6(z_{2}/6)(\Delta z/6).
\end{equation}

These formulae correctly tend to a constant value of B=1.09 in the
limit of small values of $C$ and to B directly proportional to $C$ for
high $C$ values.  Equation (\ref{BofCapproximation}) gives an excellent fit 
(better than 6\%) for
$z_{2}=6$ and all values of $\Delta z$ as shown in Figure 6
(left-hand panel), as well as for cases of $\Delta z=1$ and any value
of $z_{2}$ (right-hand panel). For all other cases, the accuracy is
still within about 12\%, which is adequate for most applications.  The
fit can be optimized for specific needs by adjusting the coefficient of
the $z_{2}\Delta z$ term in equation (\ref{exponentp}).




\clearpage


\begin{figure}
\plotone{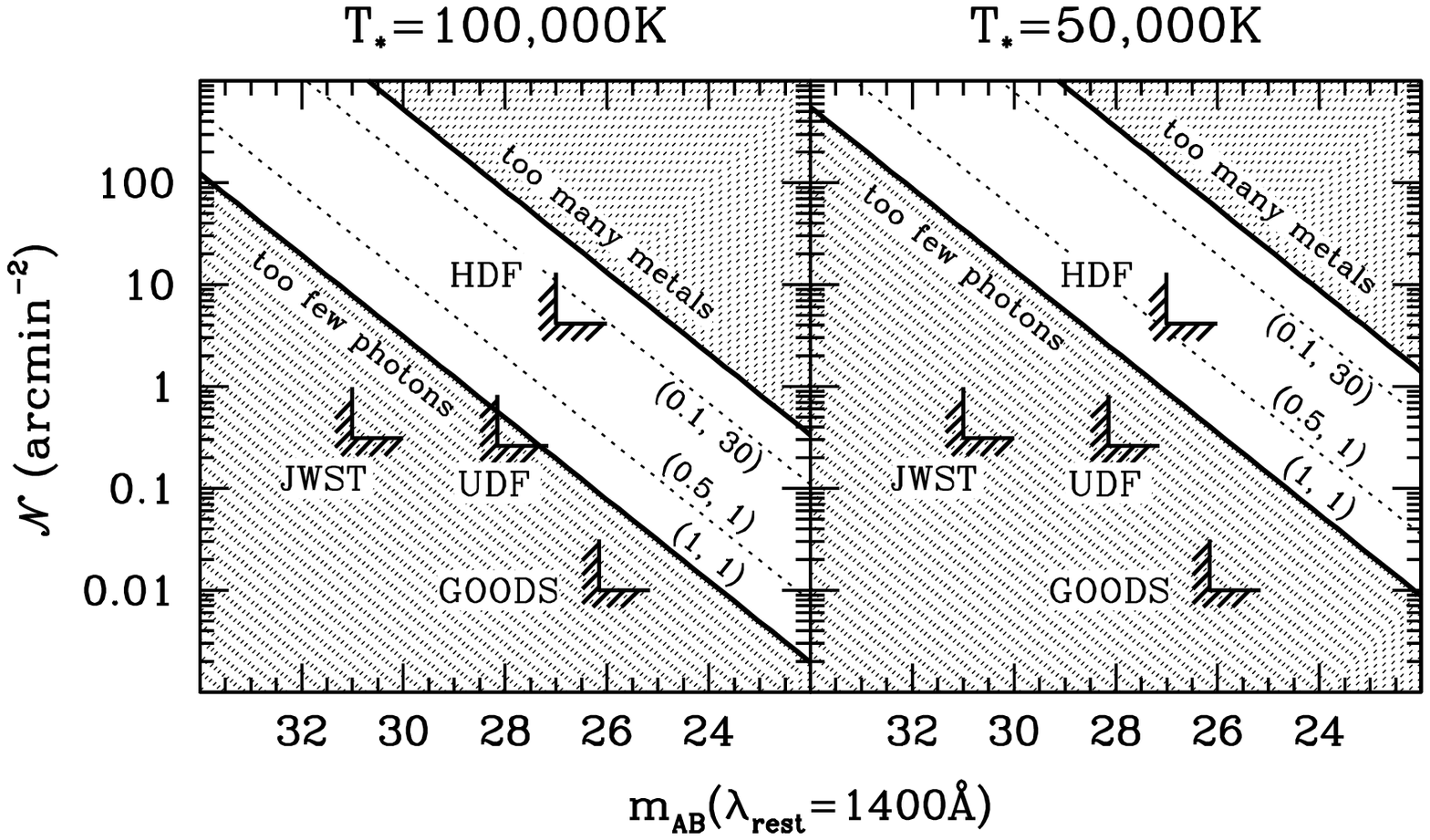}
\caption{Surface density vs apparent AB magnitude for identical reionization sources at $z_2=6$, $\Delta z=1$. 
A line represents the locus of identical sources with a given mean surface brightness.
The bottom solid line represents the minimum surface brightness model, $(f_c,C)=(1,1)$, while the upper solid line represents the
global metallicity constraint $Z\leq 0.01 Z_\odot$ at $z=z_2$.
The thin dotted lines represent the
(0.5, 1) and (0.1, 30) models which are characterized by a lower escape fraction of ionizing photons (see text). 
The non-shaded area is the only one accessible to reionization sources that do not overproduce metals.
The L-shaped markers delimit the
areas probed by the GOODS/ACS survey, the HDF/HDFS NICMOS fields, the Ultra Deep Field (UDF) and
by a hypothetical ultra-deep survey with JWST. The left panel refers to pop III sources with effective temperature of
$10^5$ K. The right panel to reionization sources with effective temperature $5 \times 10^4$ K.
\label{fig1}}
\end{figure}

\clearpage

\begin{figure}
\plotone{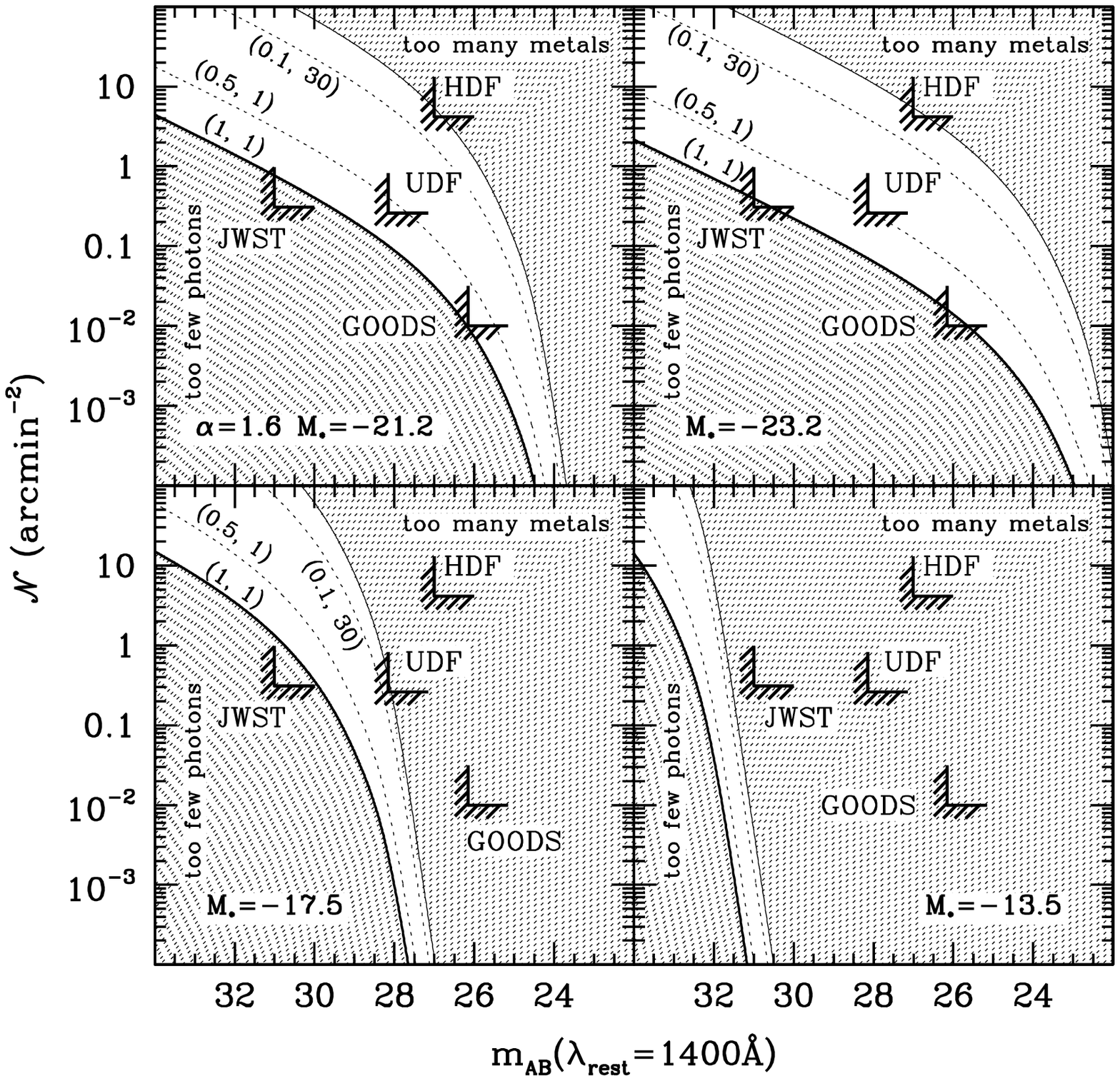}
\caption{Surface density vs apparent AB magnitude of reionization sources with luminosity
functions with different knees. The lower solid line represents the minimum surface brightness model, (1,1), for $z_2=6$ and $\Delta z=1$. The top thin solid line represents the global
metallicity constraint.
The thin dotted lines give the luminosity function for the (0.5,1) and the (0.1, 30) models (see text).
The non-shaded area is the only one accessible to reionization sources with the specified luminosity function.
The top left panel refers to the $z=3$ Lyman-break luminosity function, the remaining panels explore the effect of
changing the knee $M_{*,1400}$. 
The L-shaped markers delimit the
areas probed by the GOODS/ACS survey, the HDF and HDFS NICMOS fields, the UDF,
and an ultra-deep survey with JWST. If the value of $M_{*,1400}$ were as faint as -13.5, the reionization sources
would be hard to detect even with JWST.
\label{fig2}}
\end{figure}

\clearpage

\begin{figure}
\plotone{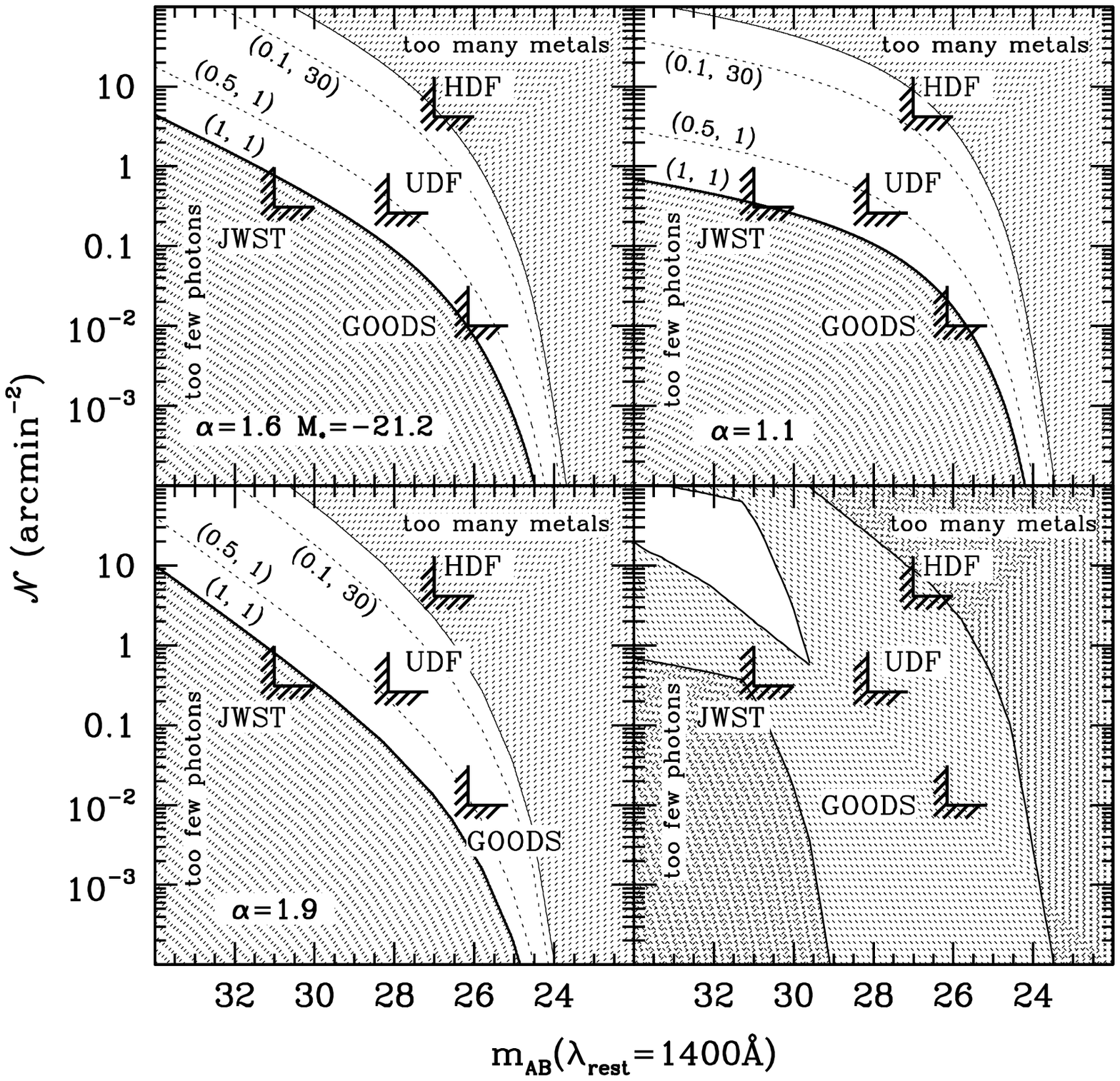}
\caption{Surface density vs apparent AB magnitude of reionization sources with luminosity
functions with different faint-end slopes. The lower solid line represents the minimum surface brightness model, (1,1), with $z_2=6$ and $\Delta z=1$. The top thin solid line represents the global
metallicity constraint.
The thin dotted lines give the luminosity function for the (0.5,1) and the (0.1, 30) models (see text).
The non-shaded area is the only one accessible to reionization sources with the specified luminosity function.
The top left panel refers to the $z=3$ Lyman-break luminosity function, the top right and bottom left panels explore the effect of changing the faint-end slope $\alpha$. The bottom right panel shows the forbidden region
(doubly shaded) and the preferred region (non-shaded), corresponding to luminosity functions with $\alpha$ ranging
between 1.1 and 1.6 and $M_{*,1400}$ ranging between -16.2 and -21.2.
The L-shaped markers delimit the
areas probed by the GOODS/ACS survey, the HDF and HDFS NICMOS fields, the UDF,
and an ultra-deep survey with JWST. 
\label{fig3}}
\end{figure}

\clearpage

\begin{figure}
\plotone{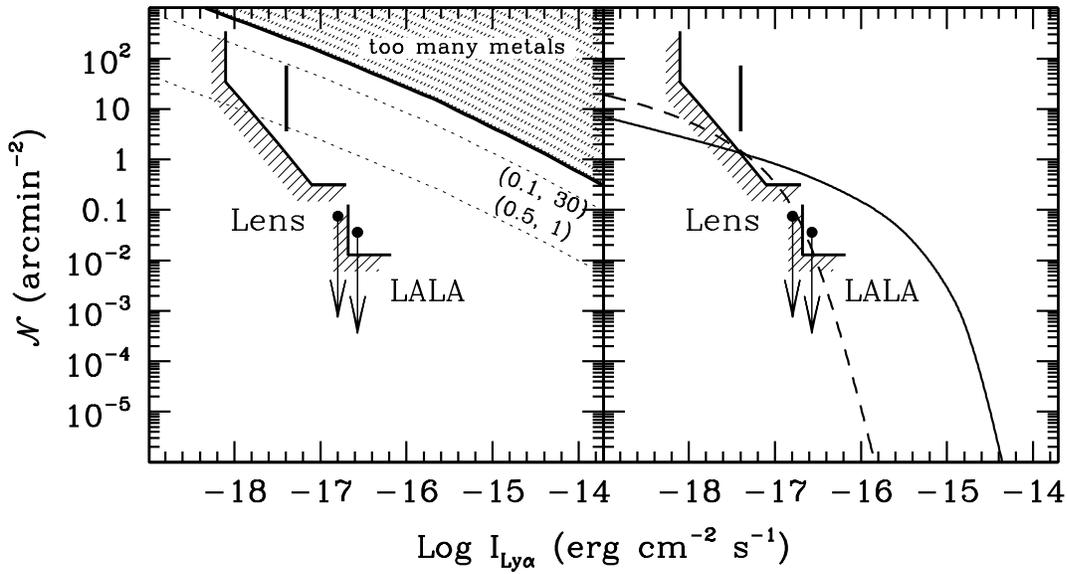}
\caption{Surface density vs Ly$\alpha$ line flux of reionization sources. 
The thin dotted lines in the left panel indicate for two different $(f_c, C)$ models the locus of  identical sources
for $z_2 = 6$ and $\Delta z=1$ (see text). 
The top solid line
identifies the global metallicity constraint. The luminosity function adopted in the figure has $\alpha=1.6$ and
$M_*=-16.2$. 
The right panels illustrates two different luminosity functions. The
solid line refer to a luminosity function identical to that of $z=3$ Lyman break galaxies, while the dashed line
refers to a luminosity function with $M_{*,1400}=-17.5$ and a slope identical to the local slope, $\alpha=1.1$.
In both panels, the L-shaped marker identifies the
area probed by a narrow-band excess survey at $z \geq 6$. The oblique marker labeled Lens represents a hypothetical 100-orbit survey with the ACS grism on a cluster of galaxies to exploit gravitational amplification.
The solid
bar represents the density estimated from the detection at $z=6.56$ by \citet{hu02} while the two points with down-pointing arrows represent their upper limits. 
\label{fig4}}
\end{figure}

\clearpage

\begin{figure}
\plotone{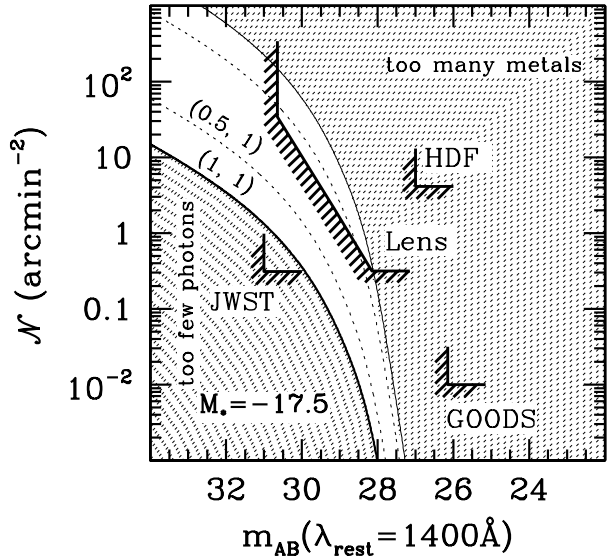}
\caption{Surface density vs apparent AB magnitude of reionization sources with a broad luminosity
function. The lower solid line represents the minimum surface brightness, (1,1), model with $z_2=6$ and $\Delta z=1$. The top thin solid line represents the global
metallicity constraint.
The thin dotted lines give the predictions for the (0.5,1) and the (0.1, 30) models (not labeled, see text).
The adopted luminosity function has faint-end slope $\alpha=-1.6$ and knee $M_{*,1400}=-17.5$.
The non-shaded area is the only one available to reionization sources with this luminosity function.
The L-shaped markers delimit the
areas probed by the GOODS/ACS survey, the HDF and HDFS NICMOS fields,
and an ultra-deep survey with JWST. The oblique marker labeled Lens shows the area probed by a 400-orbits survey with ACS pointed toward a cluster of galaxies as gravitational lens. If the value of $M_{*,1400}$ is as faint as -17.5, ACS will be able to detect the reionization sources
only if they are relatively inefficient and only if the surveys exploits gravitational amplification.
\label{fig5}}
\end{figure}

\clearpage

\begin{figure}
\plotone{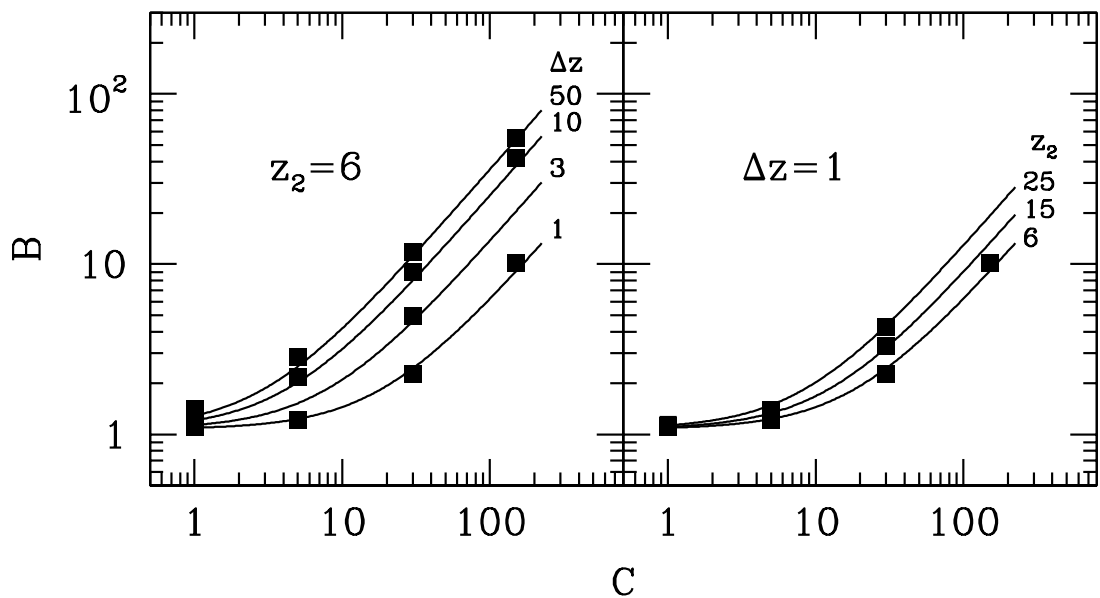}
\caption{The left panel shows the mean number of photons per ionization $B$ as a function of the IGM clumpiness factor $C$ for $z_2=6$ and a variety of values of $\Delta z$. The solid squares were obtained by direct integration of equation (8), while the lines have been obtained with equation (\ref{BofCapproximation}). The right panel shows $B$ as a function of $C$ for a given value of $\Delta z$ and different values of
$z_2$. The variations are more modest because, for a given $\Delta z$, the shorter time interval at increasing $z_2$ is compensated by the higher recombination rate due to the higher density.
\label{figA1}}
\end{figure}






\clearpage

\begin{deluxetable}{cccccccc}
\tabletypesize{\scriptsize}
\tablecaption{Mean surface brightness for reionization \label{tbl-1} for H+He IGM and $T_*=10^5$ K}
\tablewidth{0pt}
\tablehead{
\colhead{$z_2$} & \colhead{$\Delta z$}  & \colhead{ $C$ } & 
 \colhead{$B$} & \colhead{$\tau_T$} &
\colhead{$\mu_{1400(1+z2)}$} & \colhead{$\mu_{1400(1+{\bar z})}$}
& \colhead{$J_{{\rm Ly}\alpha}(f_c=0.5)$} \\
\colhead{} & \colhead{} &\colhead{} &\colhead{} & \colhead{} &
\colhead{(AB mag arcmin$^{-2}$)} & \colhead{(AB mag arcmin$^{-2}$)}
& \colhead{(erg cm$^{-2}$ s$^{-1}$ $\times 10^{16}$)}}
\startdata 
6.2&	1&	1&	1.115&	0.046&28.773 & 28.874 & 1.276	\\
6.2&	1&	30&	2.267&	0.047&28.003 & 28.103 & 2.594	\\
6.2&	3&	1&	1.158&	0.054&29.518 & 28.929 & 1.208	\\
6.2&	3&	30&	4.998&	0.062&27.930 & 27.341 & 5.216	\\
6.2&	10&	1&	1.260&	0.079&30.238 & 29.240 & 1.119	\\
6.2&	10&	30&	9.042&	0.097&28.099 & 27.101 & 8.031	\\
6.2&	20&	1&	1.335&	0.105&30.439 & 29.507 & 1.088	\\
6.2&	20&	30&	10.732&	0.128&28.176 & 27.245 & 8.748	\\
6.2&	30&	1&	1.373&	0.125&30.063 & 29.716 & 1.078	\\
6.2&	30&	30&	11.373&	0.150&27.768 & 27.420 & 8.930	\\
6.2&	40&	1&	1.397&	0.140&30.349 & 29.941 & 1.075	\\
6.2&	40&	30&	11.694&	0.166&28.043 & 27.635 & 8.999	\\
6.2&	50&	1&	1.413&	0.153&30.131 & 30.119 & 1.074	\\
6.2&	50&	30&	11.881&	0.180&27.819 & 27.807 & 9.032	\\
7.5&	1&	1&	1.117&	0.059&28.784 & 28.870 & 1.092	\\
7.5&	1&	30&	2.439&	0.061&27.937 & 28.022 & 2.281	\\
7.5&	3&	1&	1.166&	0.069&29.411 & 28.849 & 1.050	\\
7.5&	3&	30&	5.639&	0.077&27.700 & 27.138 & 5.076	\\
9&	1&	1&	1.119&	0.075&28.794 & 28.867 & 0.937	\\
9&	1&	30&	2.626&	0.078&27.868 & 27.941 & 2.198	\\
9&	3&	1&	1.175&	0.086&29.293 & 28.833 & 0.914	\\
9&	3&	30&	6.333&	0.096&27.464 & 27.004 & 4.926	\\
15&	1&	1&	1.129&	0.153&28.809 & 28.856 & 0.601	\\
15&	1&	30&	3.309&	0.157&27.642 & 27.688 & 1.760	\\
15&	3&	1&	1.206&	0.167&28.874 & 28.793 & 6.092	\\
15&	3&	30&	8.721&	0.182&26.726 & 26.644 & 4.407	\\
25&	1&	1&	1.141&	0.316&28.815 & 28.843 & 0.378	\\
25&	1&	30&	4.275&	0.322&27.381 & 27.410 & 1.415	\\
25&	3&	1&	1.246&	0.335&28.668 & 28.752 & 0.399	\\
25&	3&	30&	11.849&	0.355&26.222 & 26.306 & 3.791	\\
\enddata


\end{deluxetable}

\end{document}